# Topologically protected elastic waves in phononic metamaterials

S. Hossein Mousavi,[1] Alexander B. Khanikaev,[2,3] and Zheng Wang[1]

[1] Microelectronics Research Centre, Cockrell School of Engineering, University of Texas at Austin, Austin, TX 78758 USA

[2] Department of Physics, Queens College of The City University of New York, Queens, NY 11367, USA

[3] The Graduate Centre of The City University of New York, New York, NY 10016, USA

**Abstract**

**Topological states of quantum matter exhibit unique disorder-immune surface states protected by underlying nontrivial topological invariants of the bulk. Such immunity from backscattering makes topological surface or edge states ideal carriers for both classical and quantum information. So far, topological matters have been explored only in the realms of electronics and photonics, with limited range of bulk properties and largely immutable materials. These constraints thus impose severe performance trade-offs in experimentally realizable topologically ordered states. In contrast, phononic metamaterials not only provide access to a much wider range of material properties, but also allow temporal modulation in the non-adiabatic regime. Here, from the first-principles we demonstrate numerically the first phononic topological metamaterial in an elastic-wave analogue of the quantum spin Hall effect. A dual-scale phononic crystal slab is used to support two effective spins of phonon over a broad bandwidth, and strong spin-orbit coupling is realized by breaking spatial mirror symmetry. By preserving the spin polarization with an external load or spatial symmetry, phononic edge states are shown to be robust against scattering from discrete defects as well as disorders in the continuum. Our system opens up the possibility of realizing topological materials for phonons in both static and time-dependent regimes.**

One of the most intriguing advancements of the condensed matter physics is the discovery of a novel state of matter known as topologically ordered states, such as two-dimensional quantum Hall states, quantum spin Hall states, and three-dimensional topological insulators[1,2]. These topological orders were all first observed and realized in electronic materials, and were initially thought to be intimately linked with the Fermi-Dirac statistics of electrons. Only recently, topological orders were generalized[3–11], and observed[12–14] in bosonic systems,[15] for example, in periodic photonic media[16,17]. However, backscattering-immune edge states, the quintessential topological phenomenon, are excited in rather dissimilar ways, due to the different statistics between fermions and bosons: electron transport is driven by potential gradients or spin pumps, while photonic transport requires no such gradients. The topological protection from disorder-induced backscattering is of fundamental importance to photonic systems for three reasons: they offer an unparalleled tolerance towards defects and fabrication imperfection; the lack of feedback suppresses amplitude and phase noises in active systems; the absence of reflection reduces the overall system response from a complex multi-pass scattering to a simpler algebraic multiplication of the transfer functions of the constituent stages, thereby opening up the possibility of large-scale photonic circuits.

Phonons, classically known as elastic waves in solids, are also bosons that can similarly benefit from topologically protected transport. Indeed, unique advantages of phononic information processing, including much smaller wavelength (i.e. device footprint) and stronger phonon-phonon interaction[18,19] in comparison to photonic systems, originate from the speed of sound being orders of magnitude lower than the speed of light. The resultant slow group velocity and high density of states enhance backscattering,[20] and render phononic systems far more disorder-susceptible than photonic systems. Large contrast in acoustic impedance between common materials further promotes backscattering from disorder. Thus, realizing topological protection against even a sub-class of structural imperfections and disorders has dramatic implications for practical applications. Moreover, realizing topological orders of phononic states is of scientific interest: phonons possess three polarization variants, i.e. *three* available spin states[21], which is fundamentally different from the two spin states available to electrons and photons. Since spin plays a pivotal role in forming topological insulators, such a new spin degree of freedom may facilitate the exploration of new topological orders.

Nevertheless, to realize topologically protected transport for phonons in solids, notably chiral edge states in quantum Hall effect and helical edge states in quantum spin Hall effect (QSHE), one must overcome several nontrivial challenges associated with symmetry and degeneracy inherent to elastic materials. First, unlike electronic and photonic systems where a static magnetic field can readily break time-reversal (T) symmetry, passive elastic materials generally conserve T symmetry[22,23], thereby precluding a phononic analogue of chiral edge states in passive materials. On the other hand, the existence of helical edge states relies on two degenerate spin states both having Dirac dispersion in the absence of spin-orbit coupling, and the lack of any "magnetic" defect that can hybridize the spin states. Most solid interfaces and surfaces are well known to mix all three polarizations of elastic waves[23], essentially functioning as "magnetic" defects. Thus, to realize helical edge states, unlike electronic and photonic materials, solid phononic materials must be deliberately designed to *simultaneously* satisfy four conditions: 1) a complete bandgap for the extra spin state to prevent its excitation; 2) degenerate Dirac dispersion for the two remaining spin states; 3) gauge fields emulating spin-orbit interaction and inducing topological order; 4) protection from spin mixing between the two spin states, i.e. absence of "magnetic" defects. These demanding conditions are the principal reason that phononic topological phases have so far been predicted only in mechanical lattices of coupled rigid bodies[24–29], scalar (p-wave) acoustic resonators[30,31], or static buckling of origami structures[32,33]. It remains a challenge to realize phononic topological phases for a general monolithic solid structure that supports all three elastic-wave polarizations and is scalable to operate at GHz and beyond.

In this Letter, we demonstrate the first solid-state mechanical system with phononic topological order, with numerical experiments illustrating topologically protected helical edge phonons. This phononic analogue of quantum spin Hall effect is realized via the following steps. First, the phonon-specific challenges are resolved with the careful use of waves in a phononic crystal made from a solid membrane with properly chosen thickness and meticulously-engineered elastic anisotropy. Mirror symmetry protects symmetric and antisymmetric waves from mixing at membrane interfaces, allowing any linear combination of the two waves to be used as candidates for the spin states. The membrane thickness, elastic anisotropy, and the crystal design ensure that not only exactly two candidate spin states exist in the frequency range of interest, but also the two states are degenerate. The next step introduces a strong spin-orbit coupling by breaking the mirror symmetry, which leads to a phase transition into two-dimensional topological insulators. The final step involves truncating the phononic crystal with a "non-magnetic" boundary without spin hybridization, via either spatial symmetry or external loading.

The first step aims to create a phononic band structure emulating the electronic band structure of graphene with two uncoupled and degenerate spin states with Dirac dispersion. Consider a dual-scale phononic crystal[34] shown in Fig.1a, formed by a triangular array of air holes perforated in a slab of elastic metamaterial[34–39]. Two scales of patterning are built into this structure for different purposes: the smaller *deep*-subwavelength patterning yields extreme elastic anisotropy, and can be well characterized as an elastic non-resonant metamaterial; the larger wavelength patterning creates a graphene-like band structure for phonons. Specifically, the in-plane triangular symmetry provides Dirac dispersion[40,41] for phonons with a coincidental degeneracy at K and K′ points (Dirac point) as shown in Fig.1b. The out-of-plane mirror-reflection symmetry $\sigma_z$ is also important. With it, all phononic modes are essentially Lamb waves with modified dispersion, and can be classified by their displacement fields as either symmetric (S) or anti-symmetric (A) modes.[23] Consequently, matching the frequency and the slope (group velocity) of Dirac cones associated with a symmetric mode and an antisymmetric mode in a frequency range with no other modes is sufficient for emulating the two spin states in graphene, known to exhibit QSHE when strong spin-orbit coupling is introduced. However, symmetric modes and anti-symmetric modes generally follow drastically different dispersion relations,[23] and thus such degeneracy does not exist in most solid plates. To address this challenge, we took advantage of the exceptionally large contrast in elastic properties between solid and air inside the subwavelength perforations, and structurally tuned a highly anisotropic non-resonant metamaterial (see Supplementary Information A) to realize the desired degeneracy as seen in Fig. 1b: a four-fold degeneracy at each Dirac point with two overlaid Dirac cones from both families of modes (Fig. 1c). Note that matching the group velocity of the Dirac cones is critical, because not only this condition ensures that the S and A modes are degenerate over a broad range of frequency around the Dirac point, but more importantly it allows one to use any unitary transform of the original orthogonal basis (S and A modes) as effective spin states. Near the K-point ($k_\parallel = k_K + \delta k_\parallel$) and using the A and S modes as the basis, this system is described by a 4x4 effective Hamiltonian:

$$\widehat{H}_{A,S}(k_\parallel) = \begin{bmatrix} v_A \widehat{\boldsymbol{\sigma}}_\parallel \cdot \delta k_\parallel & 0 \\ 0 & v_S \widehat{\boldsymbol{\sigma}}_\parallel \cdot \delta k_\parallel \end{bmatrix}, \quad (1)$$

where $\widehat{\boldsymbol{\sigma}}_\parallel = [\hat{\sigma}_1, \hat{\sigma}_2]$ are the Pauli matrices of the Dirac bands subspace (see Supplementary Information B). $v_A$ and $v_S$ are the group velocities of the A and S modes, and have an identical value of $v_D$ when degeneracy is achieved. Any bulk mode can be expanded into a linear superposition of four Dirac band eigenstates described by a four-component wavefunction $|\boldsymbol{\phi}\rangle = [\phi_A^I, \phi_A^{II}, \phi_S^I, \phi_S^{II}]$, where the superscripts *I* and *II* denote the lower and upper Dirac bands respectively.

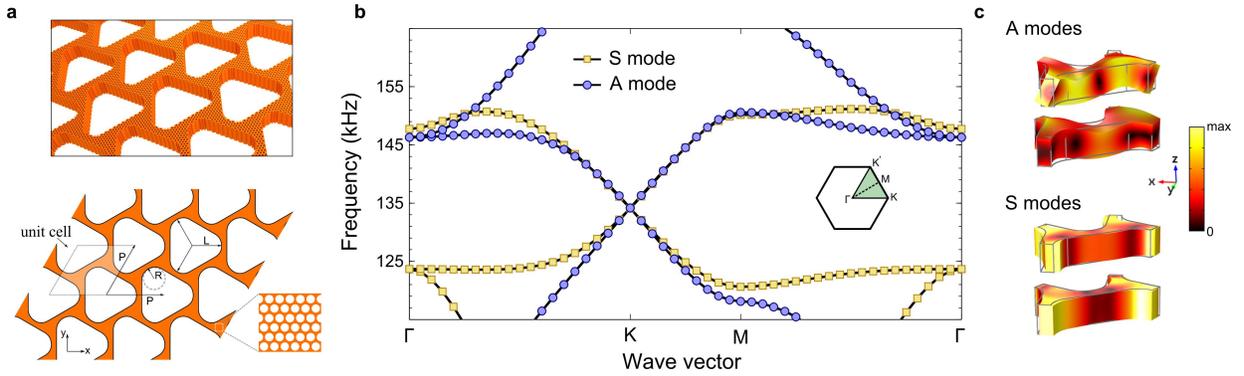

**Figure 1| Dual-scale phononic crystal with degenerate Dirac cones:** (a) Perspective view (upper panel) and top view (lower panel) of a phononic crystal made of a triangular lattice of air holes in a slab of aluminium metamaterial. P=1 cm; L=5.25 mm; R=1.95 mm. The non-resonant metamaterial slab is 2.54 mm thick, with sub-wavelength air holes at a filling ratio of 0.65. (b) Phononic band structure with degenerate Dirac points and Dirac velocities, along the irreducible Brillouin zone boundary (shown as

inset). (c) Displacement fields of a unit cell at the Dirac point (K). Colour indicates the displacement amplitudes from the undeformed configuration (grey contours).

In the second step, we introduce strong spin-orbit coupling to induce topological phase transition, accompanied by the opening of a topological band gap at the K and K′ points for the bulk crystal (Fig. 2). Note that the degeneracy between the A and S modes achieved at the previous step ensures a *complete* phononic band gap in the proximity of the former Dirac points, forming an "insulating state" for the phononic crystal (Fig. 2b).[42] The use of Lamb waves is crucial to the formation of the complete bandgap[43,44]. By enlarging the upper rim of the air holes into a counterbore structure, as shown in Fig. 2a, an effective gauge field emulating spin-orbit coupling[45] is introduced. This structural change breaks $\sigma_z$ mirror symmetry, and is designed to induce coupling within two mode pairs in the original Dirac bands: the lower $A^I$ mode and the upper $S^{II}$ mode, as well as the lower $S^I$ mode and the upper $A^{II}$ mode. For all frequencies near the original Dirac points where the frequency and group velocity degeneracy are maintained, the eigenmodes of the system become hybridized as $(A + S)/\sqrt{2}$ and $(A - S)/\sqrt{2}$, which will be used as the two effective spins. Inspection of the numerically calculated displacement fields confirms such pairwise hybridizations of the A and S modes as the new eigenmodes (Fig. 2c).

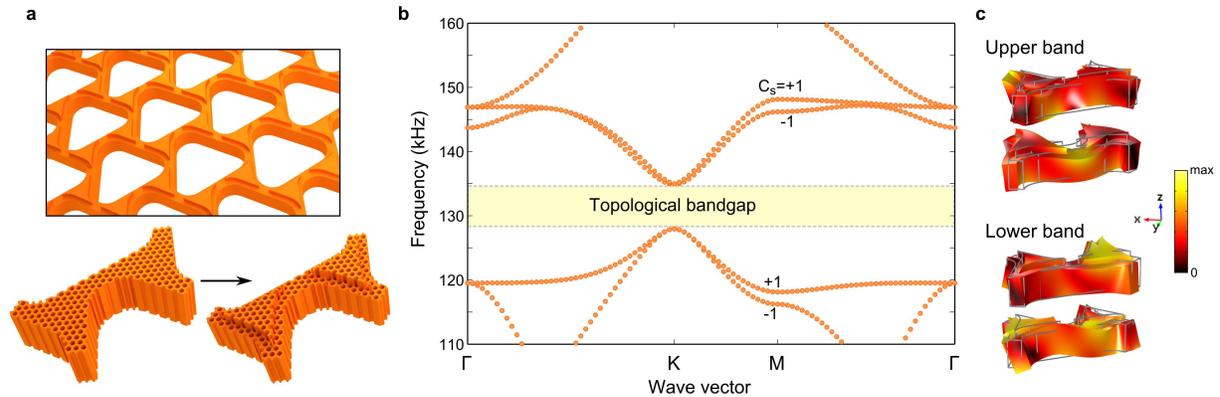

**Figure 2| Spin-orbit coupling and band gap opening by breaking mirror symmetry $\sigma_z$.** (a) Perspective view of the modified phononic crystal (upper panel) with broken z-mirror symmetry: the top rim of each air hole is enlarged into a counterbore (lower panel) with 14% increase in size and a 20% depth of the overall thickness. The overall thickness is increased to 2.94 mm to restore the spin-

degeneracy between the modes at the K point. (b) Phononic band structure showing a complete band gap (5.4% relative bandwidth) induced by the symmetry breaking. Phase and group velocities at the band edge remain matched near the K point. Spin Chern number $C_S$, calculated using first-principles FEM simulations, is shown for each band. (c) Displacement fields of a unit cell at the Dirac point (K), illustrating the hybridization between the S and A modes. Colours indicate the absolute value of the displacement.

Treating the structural modification as a perturbation, we mapped the effective Hamiltonian to the Kane-Mele theory[45], proving this system is a phononic analogue of quantum spin Hall effect. Indeed, keeping the A and S modes as the basis, the perturbation is described by a first-order correction to the unperturbed Hamiltonian $\widehat{\mathcal{H}}_{A,S} = \widehat{H}_{A,S} + \widehat{V}_{A,S}$ with $\widehat{V}_{A,S} = [0, m\hat{\sigma}_3; m\hat{\sigma}_3, 0]$, where $\hat{\sigma}_3$ is a Pauli matrix (see Supplementary Information B). Switching to the hybridized modes as the basis in the vicinity of K (K') points, we denote the hybridized eigenmodes in the presence of the effective gauge field as the + and − modes. The perturbed Hamiltonian assumes the block-diagonal form: $\widehat{\mathcal{H}}_{+/-} = [v_D \widehat{\boldsymbol{\sigma}}_\parallel \cdot \delta \boldsymbol{k}_\parallel + m\hat{\sigma}_3, 0; 0, v_D \widehat{\boldsymbol{\sigma}}_\parallel \cdot \delta \boldsymbol{k}_\parallel - m\hat{\sigma}_3]$, and is identical to the low-energy Kane-Mele Hamiltonian for quantum spin Hall effect in graphene with spin-orbit coupling:[45]

$$\widehat{\mathcal{H}}_{+/-} = v_D \hat{\tau}_0 \hat{s}_0 \widehat{\boldsymbol{\sigma}}_\parallel \cdot \delta \boldsymbol{k}_\parallel + m \hat{\tau}_3 \hat{s}_3 \hat{\sigma}_3, \qquad (2)$$

where $\hat{\tau}_i$ and $\hat{s}_i$ are inter-valley and pseudo-spin Pauli matrices. It is important to notice that because the Hamiltonian in Eq. 2 lacks the Rashba term, spin states are conserved, and spin-dependent Chern number is well defined.[46] Using first-principle finite-element method (FEM), we have numerically calculated the Berry curvature of the phononic bands immediately above and below the topological bandgap in the k-space and found the corresponding spin Chern numbers to be $C_s = \pm 1$ (see Supplementary Information C).

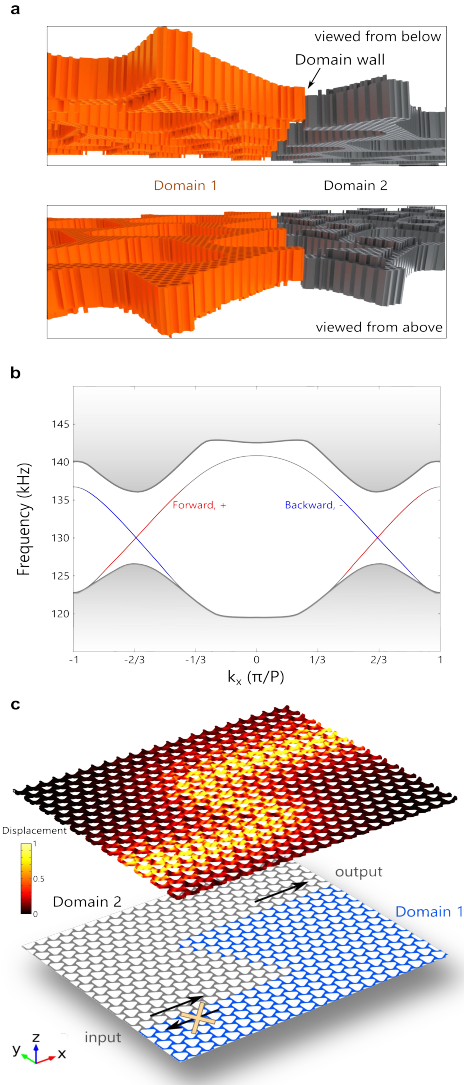

**Figure 3| Topologically protected phononic helical edge modes propagating along the domain wall:** (a) Domain wall between two crystals with the counterbore made on opposite faces, shown in two viewing angles. The sign of the effective mass is reversed across the domain wall. (b) Band structures of the phononic helical edge modes at the domain wall and the projected bulk bands (grey). The forward edge mode (red) and the backward edge mode (blue) are of the opposite spin. (c) Displacement fields of the forward edge mode demonstrating reflectionless propagation along a zigzag shaped domain wall. Colours indicate the amplitude of the displacement.

The final step toward topologically protected transport is the truncation of the bulk crystal without causing coupling of the two spin states, in order to support helical edge states. To this end, we introduce a domain wall, across which the sign of spin-orbit coupling (the term $m$ in Eq. 2) is reversed. Structurally, this domain wall is a boundary between two areas of the slab, where the counterbores are located at the opposite faces of the crystal (Fig. 3a). The bulk crystals on opposite sides of the domain wall thus possess opposite signs in their spin Chern number as well. According to the bulk-boundary correspondence principle,[47] topologically protected edge states emerge at this boundary. First-principle FEM simulations of such a structure are shown in Fig. 3. Two counter-propagating modes reside in the band gap in the place of former Dirac cones near the K point (Fig. 3b). The displacement fields of the two modes reveal that they are well localized to the domain wall (see Supplementary Movies) and carry opposite spin as predicted by the perturbation theory (see Supplementary Information B). Their time-reversed counterparts also exist near the $K'$ point and similarly carry opposite spins. The spin of all four edge states is locked to their respective propagation direction, ensuring their topological robustness. Moreover, such band structure is robust in viscous ambient: immersion in air or water only adds a miniscule attenuation that is negligible for the length scales considered here (see Supplementary Information G).

To verify the most striking feature of QSHE, i.e. the topological protection of helical edge states from back-scattering caused by spin-conserving non-magnetic disorders as defined by the gauge field, we performed a set of large-scale first-principle numerical simulations. Structurally, the non-magnetic disorders defined by the gauge field include size variations on the counterbores, as well as a broad range of domain-wall imperfections along arbitrary trajectories, including closed trajectories forming local resonances. The first example presented in Fig. 3c illustrates a zigzag domain wall functioning as a phononic waveguide with acute-angled bends. Without any structural tuning at the waveguide bends a complete transmission of the forward spin-up $|+\rangle$ edge mode can be observed and no standing wave pattern is present. This phenomenon is in sharp contrast to conventional waveguides exploiting topologically trivial surface or interface modes, for which structural modification at the waveguide bends is needed to accomplish complete transmission[48]. A more general variety of disorders, a domain wall with arbitrary turns and angles representing a one-dimensional random potential, is shown in Figure 4 to illustrate the topological robustness of the helical edge states. Such a strong random potential is known to cause localization of wave and its back-reflection in conventional waveguides without topological protection, as no discernible amount of power reaches the output in Fig. 4a. A striking difference is seen in Fig. 4b: the helical edge mode propagates with no backscattering along the path of the same random potential. It is worth emphasizing that, in comparison to photonic systems, such topological protection is of particular significance to phononic circuits, because typical acoustic impedance contrast (e.g. between aluminium and air) is many orders of magnitude greater than that in photonics. Furthermore, the need for topological protection is motivated by the inherently larger density of states for phonons, roughly $10^5$ times greater than that of photons, which causes far greater backscattering.[20,49] Also note that the bandwidth of the topological protection extends over the entire topological band gap, and similar robustness for the backward running $|-\rangle$ mode has been verified as well.

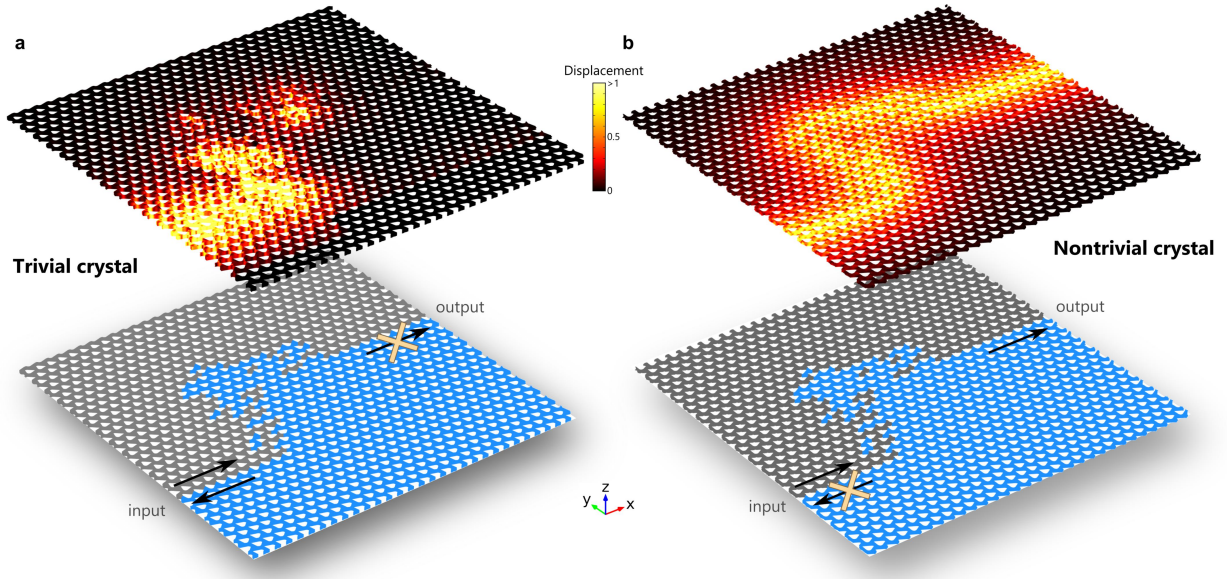

**Figure 4| Topologically trivial and nontrivial edge modes facing a one-dimensional random potential:** (a) Trivial crystal is designed using the crystal in Fig. 1 with the triangular holes rotated by 12.5 degrees clockwise (gray region) and counterclockwise (blue region). To create the bandgap in the same spectral range as in the nontrivial case, the structure is scaled up by factor of 1.07% in all

geometrical parameters. In this structure with mirror symmetry, the symmetric mode is excited. (b) The nontrivial crystal is exactly the same as in Fig. 3c. The excitation frequency in both panels is 130.5 kHz.

Replacing conventional waveguide with topologically protected helical edge states can also overcome many circuit- and system-level performance limits induced by scattering. For example, without backscattering, multi-path feedback between resonators is eliminated, and thus large numbers of resonators can be integrated in high spatial and spectral density without mutual interference. In a conventional waveguide (see Supplementary Information D), undesirable reflection occurs near the resonance of a side-coupled standing-wave resonator (Fig.5a), oftentimes as a result of mode splitting between traveling-wave resonances. With cascaded resonances, the reflection causes the overall system response to differ drastically from the product of responses of individual resonators, and also leads to long-range coupling and frequency-pulling between resonators of similar frequencies, regardless of the physical distance between the resonators. These complications hinder large-scale integration, and necessitate the use of much larger traveling-wave resonators or nonreciprocal elements. However, the lack of nonreciprocal phononic materials at frequencies beyond MHz represents a major practical challenge[50]. In stark contrast, the phononic helical edge modes, also side-coupled to a resonator, experience no reflection, with the resonant effect manifesting exclusively in the phase response (Fig.5b). Here even though the resonator is created with a closed-path domain wall with no apparent symmetry, no standing-wave resonance can be formed as long as the spin states are conserved. The complete transmission allows the phase response to be additive for cascaded systems, and a large number of resonators can be used for filtering or field enhancement without concerning the inter-resonator coupling, provided that they are moderately spaced to prevent near-field coupling. Such integration can enable significant increase in the capacity of spatially and spectrally multiplexed communication systems.

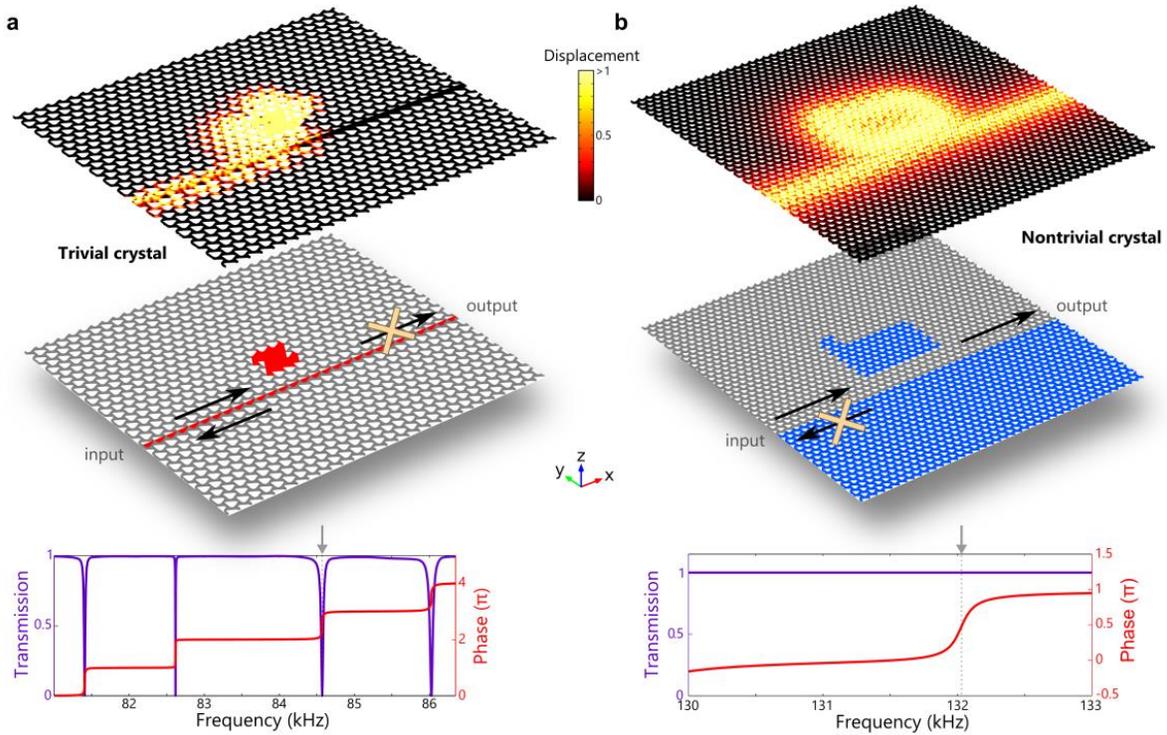

**Figure 5| Topologically-protected transmission from a resonator**. (a) Strong backscattering of a topologically trivial edge state from a coupled resonator. Complete reflection can be seen in both the displacement fields (top panel) and the transmission spectrum between the input and the output (bottom panel), when the input is tuned to the resonance (grey arrow) of a side-coupled cavity (red area in the middle panel). The edge state travels along a line defect formed by half-filled air holes (middle panel). Red areas indicate the shape of the defects, which are also filled with the phononic metamaterial (the same as grey area). The bulk of the crystal is similar to that given in Fig. 1. Arrows and crosses illustrate the direction of the *on-resonance* power flow. (b) Lack of backscattering for a phononic helical edge state, evident in the displacement fields (top panel) and the transmission spectrum (bottom panel), even at the resonance frequency of a side-coupled cavity (blue L-shaped regions in the middle panel). Both the edge state and the resonant mode are guided by domain walls between two crystals with opposite effective mass (grey and blue areas in the middle panel). The direction of the on-resonance power flow is shown in arrows. The presence of the resonator only manifests in the phase spectrum (bottom panel).

A unique advantage of phononic systems over electronic or photonic systems is the possibility of applying external loads, for example, with an array of piezoelectric actuators, to polarize helical edge states to a single spin. With the extra degree of freedom from external loads, one can truncate a bulk topological insulator and host only a single unidirectional helical edge state, reminiscent of waveguide isolators built from chiral edge states. In comparison, photonic analogues of QSHE require two domains
12

to maintain the spin polarization, while for phononic systems outlined here, one domain can be replaced by a distributed external forcing function that matches the traction created by the propagation of the desired helical edge state at the domain wall between the two crystals. Here, the uniqueness of the solution to the elastic wave equations ensures the equivalence between the replaced solid domain and the substituted external load distribution providing the identical boundary stress. Each one of the four edge modes is associated with a unique load distribution with both nontrivial normal and tangential components. In other words, a free edge (i.e., zero traction) of a topological phononic crystal does not support helical edge states, because the spin states are not conserved (see Supplementary Information E). However, applying the specified external forcing function preserves the spin state of interest, and also eliminates the other three propagating edge modes, a direct result of the orthogonality of the four helical edge states originally formed between two domains. An example of the externally-loaded truncation of the topological phononic crystal is presented in Fig. 6, with a single one-way edge mode allowed along its truncated edges. With its spatio-temporal distribution, this external load breaks time-reversal symmetry, and the associated one-way propagation can be exploited as nonreciprocal phononic devices, such as isolators. Moreover, such single mode operations are generally favoured over multimode operations in signal processing and communication systems, as intermodal interference and the subsequent phase decoherence is completely avoided.

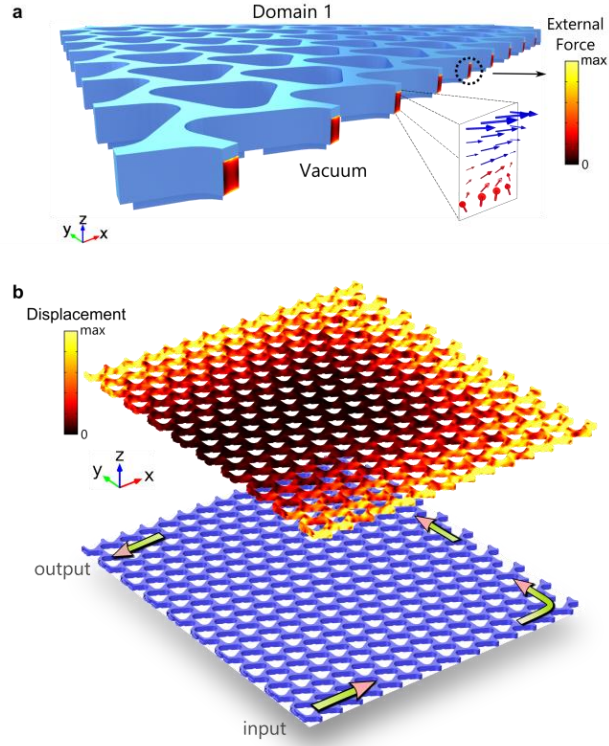

**Figure 6| Spin-polarized phononic helical edge state from external loads.** (a) External force fields on the truncated boundary of a topological crystal exposed to vacuum, which supports only the spin-up phonons traveling in the counter-clockwise fashion. The force amplitude is shown in pseudocolor, and the inset illustrates the force directions (blue for outward directions and red for inward directions) (b) Displacement fields of the resultant spin-up mode excited by a source located at the lower left corner of the crystal. Impedance-matched absorbers are used on the upper-left corner to absorb the output wave.

This work demonstrates that judicious engineering of solid structures can give rise to unusual topological systems supporting disorder-immune helical edge states for phonons. This approach opens up possibilities to realize novel topological phononic materials in both static and time-dependent regimes. Although our discussion focuses on frequencies near 100 kHz, the scale invariance of the elastic wave equation allows the design procedure to be readily scaled to higher frequencies. For example, reducing the lattice constant of the phononic crystal to 1 micron and proportionally reducing all other dimensions by 10,000 folds will raise the operational frequency to approximately 1.3GHz. Additionally, unlike photons, phonons enjoy a 100% reflection at solid/vacuum interfaces for all frequencies, further allowing this two-dimensional design to be scaled to GHz and even THz regimes. Coupled with the single mode operation using external forcing function, phononic helical edge modes represent an intriguing solution to the unfulfilled need of nonreciprocal elastic wave devices, and open up venues to explore new forms of topological orders.

**Acknowledgments**

This work is in part supported by the Packard Fellowships for Science and Engineering, the Alfred P. Sloan Research Fellowship, and Jack Kilby/Texas Instruments Endowment.

**Author contributions**

All authors contributed extensively to the work presented in this paper. S.H.M. designed the phononic crystal and performed the numerical simulations.

**Additional information**

Supplementary information is available in the online version of the paper. Reprints and permissions information is available online at www.nature.com/reprints.

Correspondence and requests for materials should be addressed to Z.W. or S.H.M.

**Competing financial interests**

The authors declare no competing financial interests.